\documentclass[twocolumn,a4,showpacs,preprintnumbers,prb]{revtex4}
\usepackage{graphicx}%
\usepackage{dcolumn}
\usepackage{amsmath}
\usepackage{amssymb}
\usepackage{bm}

\begin{document}
\title{M\"{o}ssbauer study of Na$_{0.82}$CoO$_2$ (doped by 1\% $^{57}$Fe). }
\author{M. Pissas{\protect \cite{cor}}, V. Psycharis, D. Stamopoulos, G. Papavassiliou, Y. Sanakis, A.
Simopoulos} \affiliation{Institute of Materials Science, NCSR,
Demokritos, 15310 Ag. Paraskevi, Athens, Greece}
\date{\today }
\begin{abstract}
We studied by M\"{o}ssbauer spectroscopy  the Na$_{0.82}$CoO$_2$
compound using  1\% $^{57}$Fe as a local probe which substitutes
for the Co ions. M\"{o}ssbauer spectra at $T=300$ K revealed two
sites which correspond to Fe$^{3+}$ and Fe$^{4+}$. The existence
of two distinct values of the quadrupole splitting instead of a
continuous distribution should be related with the charge ordering
of Co$^{+3}$, Co$^{+4}$ ions and ion ordering of Na(1) and Na(2).
Below $T=10$ K part of the spectrum area, corresponding to
Fe$^{4+}$ and all of Fe$^{3+}$, displays broad magnetically split
spectra arising either from short-range magnetic correlations or
from slow electronic spin relaxation.
\end{abstract}
 \maketitle

Recently, it has been discovered\cite{takada03} that H$_2$O
intercalation in the Na$_x$CoO$_2$ compound induces
superconductivity with $T_c\sim 5$ K, triggering a large research
activity. Besides this remarkable discovery, the understanding of
the magnetic properties of the triangular geometrically frustrated
CoO$_2$ plane is a very important topic in strongly correlated
electron systems. Novel types of magnetic transitions and ground
states are expected due to the suppression of a long range
magnetic ordering by the geometrical frustrations.

\begin{figure}[h] \centering%
\includegraphics[angle=0,width=0.8\columnwidth]{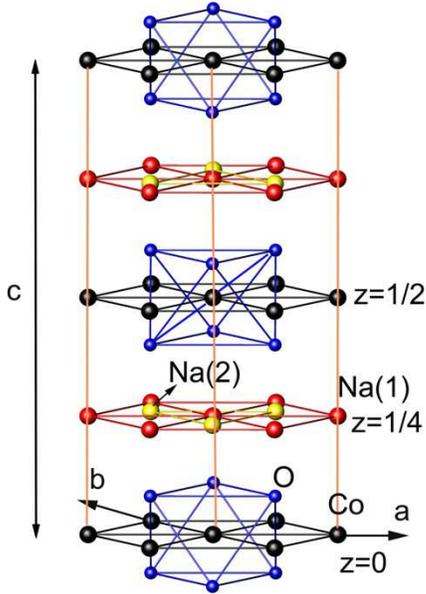}
\caption{Crystal structure of the Na$_{0.82}$CoO$_2$ compound. }
\label{fig1}%
\end{figure}%
\begin{figure}[btph] \centering%
\includegraphics[angle=0,width=0.8 \columnwidth]{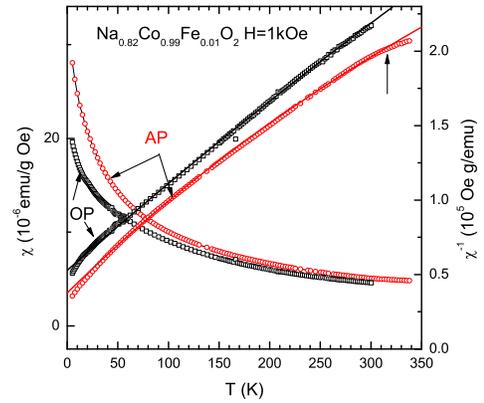}
\caption{Temperature variation of the mass susceptibility $\chi$
(left axis) and $\chi^{-1}$ (right axis), of the AP and OP
Na$_{0.82}$CoO$_2$ samples. The symbols represent the experimental
points, while the solid lines, are plots of the inverse
susceptibility $\chi^{-1}$ using the parameters mentioned in the
text. The vertical arrow shows the temperature where deviation
from the linearity is observed.}
\label{fig2}%
\end{figure}%

In addition, this category of compounds has been used as cathodic
material for solid state batteries\cite{tarascon01} and
thermoelectric devices.\cite{terasaki97} The basic characteristic
of these compounds is the non-stoichiometry which permits
reversible incorporation of foreign atoms in their lattice,
without modification of the local configuration of the atoms or
crystallographic structure. Depending on $x$, Na$_x$CoO$_2$
compound displays several crystal structures.\cite{fouassier73}
The crystal structure is in general hexagonal, consisting of {\it
edge}-sharing CoO$_6$ octahedra in between which the sodium ions
are intercalated within a trigonal prismatic or octahedral
environment. The CoO$_2$ forms a 2D hexagonal layer structure (see
Fig. \ref{fig1}). Due to the deficiency at Na site $1-x$ holes are
doped into the band insulating state of low spin Co$^{3+}$ ($S=0$,
$3d^6$ in $t_{2g}$ orbital). Alternatively, one can consider that
the Na deficiency is equivalent with $x$ electron doping in the
triangular lattice consisting of Co$^{4+}$ $S=1/2$. Besides for
the elucidation of the mechanism which is responsible for the
superconductivity, there are still several open questions
concerning the charge
segregation\cite{ray99,mukhamedshin04,ning04,bernhard04} in
CoO$_2$ layers in the samples with $x\sim 3/4$, asking for further
study. In the present paper we study the physical properties of
the Na$_x$CoO$_2$ compound at a local level, using as local probe
the iron impurity, which substitutes for the cobalt ions.

A sample with nominal composition
Na$_{0.82}$Co$_{0.99}$Fe$_{0.01}$O$_2$ was prepared by solid state
reaction of Na$_2$CO$_3$, Co$_3$O$_4$ and $^{57}$Fe$_2$O$_3$ at
$T=800^{\rm o}$C for 24h in air. Subsequently, the initial sample
was separated into two parts at which we applied different heat
treatment. The first part was additionally heated in air
atmosphere at $T=800^{\rm o}$C, for  24 h. We call this sample AP.
The second part was annealed at $T=800^{\rm o}$C for  24 h in
oxygen atmosphere and we call this OP sample. X-ray powder
diffraction (XRD) data were collected with a D500 SIEMENS
diffractometer, using CuK$\alpha $ radiation and a graphite
crystal monochromator, from 4$^{\text{o}}$ to 100$^{\text{o}}$ in
steps of 0.03$^{\text{o}}$ in $2\Theta $. The power conditions
were set at 40KV/35mA. The aperture slit as well as the soller
slit were set at $1^{\text{o}}$. The absorption M\"{o}ssbauer
spectra (MS) were recorded using a conventional constant
acceleration spectrometer with a $^{57}$Co(Rh) source moving at
room temperature, while the absorber was kept fixed in a variable
temperature cryostat. The resolution was determined to be $\Gamma
/2=0.12$ mm/sec using a thin $\alpha $-Fe foil. DC magnetization
measurements were performed in a SQUID magnetometer (Quantum
Design). $^{23}$Na NMR line shape measurements of the central
transition $(-1/2\rightarrow 1/2)$ were performed on a home made
spectrometer operating in external magnetic field ${\cal H}=8$,
Tesla. The spectra were obtained from the Fourier transform of
half of the echo. The spin echo was generated with a two-pulse
$\pi /2$-$\tau $-$\pi $ spin-echo pulse sequence.

Fig.\ref{fig2} shows the temperature variation of the mass
magnetic susceptibility ($\chi$) and the inverse magnetic
susceptibility ($\chi^{-1}$) for the AP and OP samples. The
inverse magnetic susceptibility displays a nearly linear variation
with temperature, implying a paramagnetic behavior with a small
temperature independent magnetic susceptibility. Deviation from
the Curie-Weiss equation was found below $\sim 36$ K and $\sim 15$
K for AP and OP samples respectively, suggesting that magnetic
correlations take place below those temperatures. An additional
deviation from the linearity of the $\chi^{-1}$ curve, was
observed for the AP sample at $T\approx 315$ K (see arrow in Fig.
\ref{fig2}) most probably being related with the
$H1-H2$-structural transition.\cite{huang05} Thus, we fitted the
experimental data for $36<T<315$ K for AP sample and for $T>15$ K
for OP sample, using a nonlinear least-square method, using the
Curie-Weiss formula $\chi=C/(T-\Theta)+\chi_{\rm o}$, where $C$ is
the Curie constant, $\Theta$ is Weiss temperature, and $\chi_{\rm
0}$ the temperature independent susceptibility. The estimated from
least-square fitting parameters are: $\Theta({\rm AP})=-53(5)$ K,
$\Theta({\rm OP})=-86(1)$ K, $\mu_{\rm eff}^2({\rm
AP})=1.14\mu_{\rm B}^2$, $\mu_{\rm eff}^2({\rm OP})= 1.21\mu_{\rm
B}^2$ and $\chi_{\rm o}({\rm AP})=1.3(3)\times 10^{-6}$ emu/g Oe,
$\chi_{\rm o}({\rm OP})=4.5(4)\times 10^{-7}$ emu/g Oe for AP and
OP samples, respectively. The negative value of Weiss temperature,
for both samples, suggests that the spins interact predominantly
antiferromagnetically.
The observed effective magnetic moment $\mu_{\rm eff}$ is given by
the relationship  $\mu_{\rm eff}^2=\sum_{i=1}^4 g_i^2 x_i
S_i(S_i+1)$, where $g_i=2$ is the Land\'{e} factor, and $x_i$ and
$S_i$ is the percentage and the spin of each individual ion $i$
respectively. In the present case Co is present in the
Co$^{3+}$($S=0$) and Co$^{4+}$ ($S=1/2$) states. The M\"{o}ssbauer
spectroscopic studies to be presented below suggest that Fe is in
Fe$^{3+}$($S=\frac{5}{2}$) and Fe$^{4+}$ ($S=2$) forms with a
ratio $40/60$. Taking this into consideration, the observed
$\mu_{\rm eff}$ can be accounted for by mixed valence
configurations consisting of 37\% Co$^{+4}$, ($S_1=\frac{1}{2}$),
62\% Co$^{+3}$ ($S_2=0$) for AP sample and 39\% Co$^{+4}$,
($S_1=\frac{1}{2}$), 60\% Co$^{+3}$ ($S_2=0$) for OP sample. The
estimated magnetic moments are in good agreement with those from
other works.\cite{gavilano04,motohashi03,ray99,tojo02}

\begin{figure*}[htbp] \centering%
\includegraphics[angle=0,width=1.5\columnwidth]{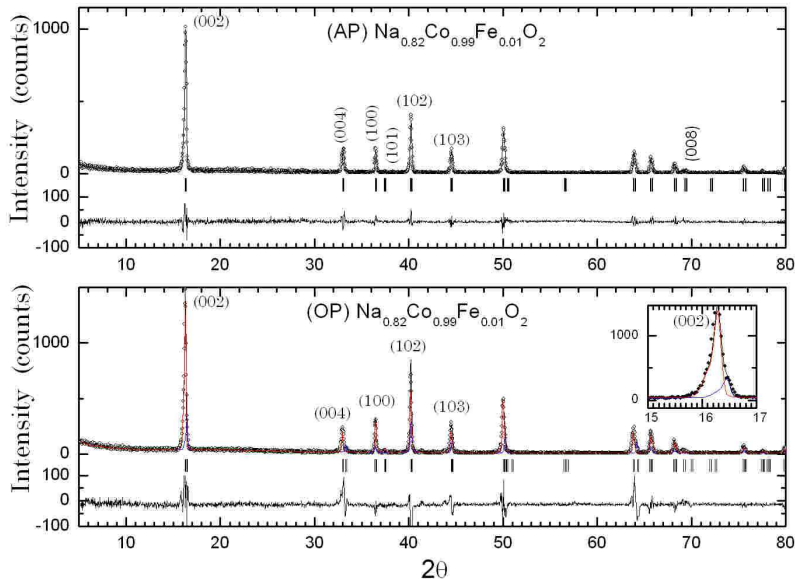}
\caption{Rietveld plots for AP and OP Na$_{0.82}$CoO$_2$ samples.
The observed data points are indicated by open circles, while the
calculated and difference patterns are shown by solid lines. The
positions of the reflections are indicated by vertical lines below
the patterns. The inset shows a zoom of the $(002)$ Bragg peak for
the OP sample. The shoulder at the high angle place correspond to
a second Na$_x$CoO$_2$ phase with slight lower $c$-axis.}
\label{fig3}%
\end{figure*}%
The x-ray diffraction data at $T=300$ K were analyzed using the
Rietveld refinement method, with the {\sc FULLPROF} suite of
programs\cite{fullprof}, assuming the hexagonal $P 6_3/mmc $ (no
194) space group for both samples. The crystal model used in the
refinement is the so called H2-structure where, Co occupies the
$2a$ $(0,0,0)$ position, O the $4f$ $(1/3,2/3,z)$, Na(1) the $2b$
$(0,0,1/4)$ while for Na(2) we tested either $2d$ $(2/3, 1/3,
1/4)$ or $2c$ $(1/3,2/3, 1/4)$ sites. The estimated unit cell
parameters were $a=2.8399(1)$\AA,~ $c=10.8301(6)$\AA~ for the AP
sample and $a=2.8382(1)$\AA,~ $c=10.8572(3)$\AA~ for the OP
sample. Fig. \ref{fig3} shows Rietveld plots for both AP and OP
samples. From very initial refinement steps we ascertained that
the Na(2) atoms exclusively occupy the $2d$ $(2/3, 1/3, 1/4)$ site
. That is, they do not "prefer" oxygen ions above and below (see
Fig. \ref{fig1}). Let us discuss firstly the AP sample.
 Firstly, we would like to point out that our
Rietveld refinement results for AP sample do not show evidence for
two phase behavior at 300 K.
After few refinement steps we achieved good agreement indexes.
Looking on refinement results we found that both Na sites are
partially occupied and display, large temperature factors $\sim 3
{\rm \AA}^2$, in agreement with the results of the Refs.
\onlinecite{balsys96,huang04}. The intensity of the $(101)$ Bragg
peak is representative of the overall Na content. The refinement
converge when 22\% of Na(1) site is occupied and 51\% of Na(2) ,
giving an overall stoichiometry Na$_{0.73\pm 0.03}$CoO$_2$ for the
AP sample. This stoichiometry is slightly lower than the nominal
one. The lower Na content is possibly related  with both the
volatility of Na oxides and un-reacted Na$_2$CO$_3$, which
amounted to a few per cent of the total. The large temperature
factors of Na sites may imply that the Na(2) atom is not located
exactly at $2d$ site. A common practice in overcoming this problem
is to permit the $x$ and $y$ coordinates of Na(2) atoms to vary
during the refinement as the $6h$ ($2x, x,1/4$), $x=0.281(4)$,
site imposes.\cite{jorgensen03} Using this model the refinement
converged at lower agreement indexes with more reasonable
temperature factors.
As far as the OP sample is concerned, the
refinement gave similar results with the AP sample. However, close
inspection of the high angle part of the (00l) Bragg peaks
revealed small shoulders (see the inset of the lower part of Fig.
\ref{fig3}). These shoulders possibly originate from crystallites
with lower $c$-axis length in comparison with the majority phase.
Based on this, we employ a two phase refinement in agreement with
Ref. \onlinecite{huang05} where their Na$_{0.75}$CoO$_2$ compound,
(which prepared in oxygen atmosphere), below $T\approx 340$ K,
displayed two-phase behavior.
This model gave substantially better agreement indexes.

\begin{figure}[ht] \centering%
\includegraphics[angle=0,width=0.8\columnwidth]{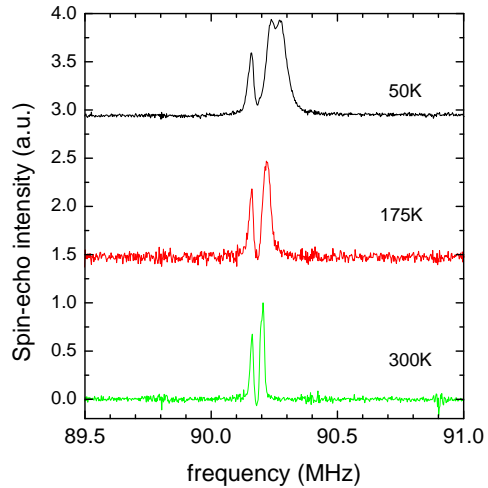}
\caption{$^{23}$Na NMR spectra at $T=50, 175$ and 300
K.}\label{fig4}%
\end{figure}

$^{23}$Na NMR spectra obtained at $T=300, 175$ and 50 K are
displayed in Fig. \ref{fig4}.
The spectra seen in Fig. \ref{fig4} concern the central part of
the spectrum, corresponding to the $-\frac{1}{2}\rightarrow
\frac{1}{2}$ transition, of our AP sample. The echo intensity,
arising from the quadrupole wings of the transitions
$\pm\frac{3}{2}\leftrightarrow\pm\frac{1}{2}$, was not observed,
probably  because our sample is polycrystalline. The spectrum, at
50 K comprises three peaks. The low-frequency line most probably
arises from un-reacted Na and/or defect Na sites. The remaining
doublet may tentatively be assigned into two Na sites, in
agreement with crystal structure data. Using similar arguments
like those of Ref. \onlinecite{mukhamedshin04} we can claim that
at low temperatures some kind of ordering in the two Na sites,
occurs. At higher temperatures the thermal motion of Na$^{+}$ ions
results in motional narrowing, causing a gradually merging of the
two lines into a single line.\cite{mukhamedshin04,gavilano04}

\begin{figure}[htbp] \centering%
\includegraphics[angle=0,width=1.0 \columnwidth]{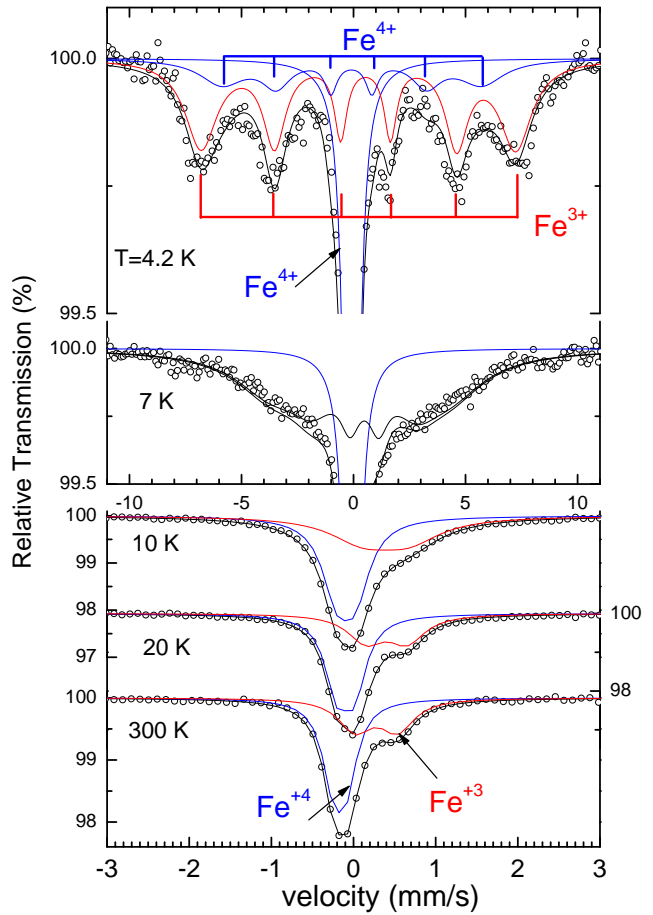}
\caption{M\"{o}ssbauer spectra of AP Na$_{0.82}$CoO$_2$ (doped by
1\% Fe-57) sample, at $T=4.2, 7, 10, 20$ and
300 K.}\label{fig5}%
\end{figure}

M\"{o}ssbauer spectra were taken for both samples at temperatures
between $T=4.2$ K and $T=300$ K. The paramagnetic and the
magnetically split M\"{o}ssbauer spectra at 300 K and 4.2 K are
least-squares fitted with two and three components, respectively
and the results are summarized in Table \ref{table2}. The spectra
of the AP sample, at $T=300$ K, consist of two components (see
Fig. \ref{fig5}) with isomer shift (with respect to Fe metal at
300 K) of -0.06 mm/s and 0.38 mm/s respectively. Based on isomer
shift values, the first component can be assigned either to low
spin Fe$^{2+}$ ($S=0$) or to high spin Fe$^{4+}$ ($S=2$).  The
second component is probably high spin ferric ($S=5/2$). Their
spectra area ratio is 60\% to 40\%. This ratio remains constant
down to $10$ K indicating that they have similar Debye-Waller
factors.
Considering that the parent compound comprises Co$^{3+}$ and
Co$^{4+}$ ions the site with small isomer shift is attributed to
Fe$^{4+}$ ($S=2$). Our assignment will be made clearer below where
we will discuss the magnetically split spectra. According to our
Rietveld results, the Na content is 0.73, implying a
Co$^{4+}$:Co$^{3+}\approx 0.25:0.75$ and as a consequence, if iron
follows the Co ions, we would expect a Fe$^{4+}$:Fe$^{3+}\approx
0.25:0.75$.  The observed area ratio of the two components is
different than the expected if we assume that the Fe impurities
substitute for Co$^{4+}$ and Co$^{3+}$ in the lattice.

Regarding the quadrupole interaction we note that the Fe$^{4+}$
component displays a small quadrupole splitting of $\sim0.18$ mm/s
and the Fe$^{3+}$ a splitting of $\sim 0.50$ mm/s. Both components
appear with narrow linewidths (0.15 mm/s for Fe$^{4+}$ and 0.24
mm/s for Fe$^{3+}$). Interestingly, although the iron randomly
substitutes the Co ions, the two iron sites, except for the
different isomer shift, have different quadrupole splitting. The
quadrupole splitting of the M\"{o}ssbauer spectra ($\Delta E_{\rm
Q}= (1/2)|e|V_{zz}Q(1+\eta^{2}/3)^{1/2}$) is directly related with
the principal value ($V_{zz}$) of the electric field gradient
tensor at the Fe nucleus. The principal value of the electric
field gradient embraces contributions from both the valence
electrons of the atom and from surrounding ions in the lattice. By
taking these contributions separately we can write
$V_{zz}/|e|=q=(1-R)q_{\rm ion}+(1-\gamma_{\infty})q_{\rm latt}$
where $R$ and $\gamma_{\infty}$ represent the effects of shielding
and antishielding respectively of the nucleus by the core
electrons.\cite{greenwood,ingalls64} For the high spin Fe$^{3+}$
due to the spherical symmetry only the lattice contribution is
significant. On the other hand, since the coordination octahedron
of the Fe$^{4+}$ ion is regular we claim that its valence
contribution in $V_{zz}$ is very low, thereby the observed
quadrupole splitting mainly arises from the lattice part too.
Consequently, it is reasonable for one to ask, why different
quadrupole splitting for the two sites is observed. Of course, the
iron as a local probe "sees" a statistical summation (not an
averaging) of all possible spectra coming from the different
configurations due to deficient Na site. A simple point charge
calculation based on different configurations of sites Na(1) and
Na(2)  demonstrate that in such a case a broad quadrupole
distribution is expected. Contrary to this we observed two
distinct values of quadrupole splitting for each valence state. In
order to explain this experimental fact we propose that a charge
ordering which include the Co$^{+3}$ and Co$^{+4}$ and/or Na ions
in Na(1) and Na(2) sites takes place producing, therefore, two
distinct environments. A similar cobalt-charge and Na ordered
configuration has also been proposed by Mukhamedshin et al.,
\cite{mukhamedshin04} Ning et al., \cite{ning04} based on their
NMR results and by Bernhard et al. \cite{bernhard04} by
interpreting their ellipsometry data, for an $x=3/4$ sample.

\begin{table}[tbp] \centering%
\caption{ M\"{o}ssbauer parameters for AP Na$_{0.82}$CoO$_2$ at
$T=300$ K and 4.2 K. Half linewidth $ \Gamma /2 $ (mm/s), isomer
shift $\delta $ relative to metallic Fe at RT (mm/s), hyperfine
magnetic field $H$ (kG), Quadrupole splitting $2\epsilon\equiv
\Delta E_{\rm Q}$, [$\epsilon =
(1/4)|e|V_{zz}Q(1+\eta^{2}/3)^{1/2}$, and
$\epsilon=(|e|V_{zz}Q/8)(3\cos^2\Theta-1+\eta\sin^2\Theta \cos
2\Phi)$ for the paramagnetic, and the magnetic (first order
perturbation theory) cases, respectively]. The numbers in
parentheses are estimated standard deviations referring to the
last significant digit.}
\begin{ruledtabular}
\begin{tabular}{cccccccccc}
\multicolumn{4}{c}{$T=4.2$ K}\\
    &SITE I(a)  & SITE II & SITE I(b) \\
    &Fe$^{4+}$  & Fe$^{3+}$ & Fe$^{4+}$ \\
\tableline
$\Gamma/2$      & 0.230(2) & 0.24    & 0.16 \\
$\delta$      & 0.017(2) & 0.484(5)& 0.020(1)\\
$\epsilon$    & 0.10(1)  &-0.123(5)& 0.01(5)\\
$H$           & 0        & 438(4)  & 356(8) \\
$\Delta H$    & 0        & 37(3)   & 61(5) \\
Area          & 40(1)    & 43(1)   & 17(5)\\
\tableline
\multicolumn{4}{c}{$T=300$ K}\\
              &SITE I    & SITE II & \\
\tableline
$\Gamma/2$      & 0.153(7)  & 0.24(1) & -\\
$\delta$      & -0.06(1)  & 0.38(1) & -\\
$\epsilon$    & 0.09(1)   & 0.25(1) & -\\
Area          & 60(2)0   & 40(2)    & -\\
\end{tabular}
\end{ruledtabular}\label{table2}
\end{table}%

It is interesting to mention here the hyperfine parameters of the
$\alpha-$NaFeO$_2$ compound\cite{ichida70} which has similar
crystal structure ($R\bar{3}m$) and the Fe is octahedrally
coordinated with oxygen. At $4.2$ K, NaFeO$_2$ displays a
magnetically split M\"{o}ssbauer spectrum with $H=455$ kOe. At
$T=300$ K the M\"{o}ssbauer spectrum consist of a doublet with
$\delta=0.33$ mm/s and $\Delta E_Q=0.46$ mm/s. These values are
very close to those of the second site, observed in AP
Na$_{0.82}$CoO$_2$ compound, justifying the assignment to
Fe$^{3+}$. Furthermore, deintercalated Na$_x$FeO$_2$ compound
displays\cite{takeda94} two doublets, the first corresponding to
Fe$^{4+}$ and the second to Fe$^{3+}$. Contrary, however, to the
Na$_{0.82}$CoO$_2$ compound the quadrupole splitting of the
Fe$^{4+}$ component is comparable with that of Fe$^{3+}$. This
fact further supports our claim for cobalt-charge and Na ordering.
Lowering the temperature, the M\"{o}ssbauer spectra do not change
down to 10 K with a constant area ratio of $60:40$ for the
Fe$^{4+}$ and Fe$^{3+}$ components respectively. At 10 K part of
the spectrum is broadened and with further lowering of the
temperature magnetic hyperfine splitting appears together with an
unsplit peak, corresponding to the Fe$^{4+}$ component (see
Fig.\ref{fig5}). The area ratio of these two components is
inverted now to the value $40:60$, a fact that leads us to
conclude that the magnetic component arises from all the Fe$^{3+}$
paramagnetic spectra area (40\%) and part (20\%) of the Fe$^{4+}$
one. In view of this, we have least-squares fitted the 4.2 K
spectrum with two magnetic components and one unresolved doublet.
Both magnetically split components display inhomogeneous line
broadening accounted for by the $\Delta H$ parameter which
represents the full width at the half maximum of a Lorentzian
distribution of the hyperfine magnetic field. This
phenomenological parameter arises either from a static
distribution of the electronic spin $<S>$ or from spin
fluctuation. Its value decreases with temperature for spectra
taken between 10 K and 4.2 K indicating thus spin fluctuations.
Such spin fluctuations can occur in isolated paramagnetic ions
(paramagnetic spin relaxation) or in low dimension magnetically
ordered systems, which is probably the case in the present
structure. It is not clear to us why $2/3$ of the Fe$^{4+}$
component remains magnetically unsplit down to 4.2 K. We could
speculate that the charge order model is not perfect and there are
regions in the lattice that are disordered or different kind of
ordering occurs.

Another characteristic of the hyperfine magnetic field value for
the Fe$^{3+}$ component is that it is lower than the expected for
typical octahedral coordinated Fe$^{3+}$O$^{2-}_6$, ($S=5/2$). The
lower hyperfine field of Fe$^{3+}$ component can be attributed
either to covalency effects or to low dimensional  effects. Recent
NMR studies\cite{mukhamedshin04} on a sample with $x=0.66$ and our
NMR results did not display any anomaly either in the linewidth or
in the signal intensity, indicating that no magnetic transition
occurs down to 1.5 K, in agreement with our magnetization
measurements.

The M\"{o}ssbauer spectra of the OP sample are similar with the AP
sample with the exception of the area ratio of the two components
that now became 65\% to 35\% indicating that part of the Fe$^{3+}$
component was further oxidized to Fe$^{4+}$. Probably this is
related with the two phase behavior revealed from our x-ray
diffraction results for this particular sample. The 4.2 K magnetic
spectra were analyzed with the same model as for the AP sample and
the analysis gave the same hyperfine parameters as for the AP
sample. It is useful to compare the hyperfine parameters of the
Fe$^{4+}$ site with the M\"ossbauer parameters of the compound
SrFe$^{4+}$O$_3$ containing iron in Fe$^{4+}$ high spin
state.\cite{gallagher64} This compound shows a single line at room
temperature with an isomer shift of 0.054 mm/s and a magnetic
hyperfine pattern at $4.2$ K with an isomer shift 0.146 mm/s and a
$H=331$ kG. The $H$ originates from an octahedral
$d^4\,(t_{2g}^{}\;^3e_g^1,\;S=2)$ high-spin Fe$^{4+}$
configuration with some degree of covalency, bearing in mind that
$d^5\;(S=5/2)$ high-spin Fe$^{3+}$ ions can show flux densities as
low as $450 $ kG. Comparing these hyperfine parameters  with those
of site (I), there is no doubt that this site corresponds to high
spin ($S=2$), Fe$^{4+}$.

Finally, it is necessary to compare our data with those of single
crystals. The properties of samples Na$_x$CoO$_2$ ($x\sim 0.75$)
are sensitive to the details of sample preparation.
Polycrystalline powder and single crystals with the same nominal
compositions exhibit qualitatively different properties. The
single crystals prepared by the floating zone method exhibit a
antiferromagnetic transition at 22 K, and a sharp first order
transition at about 340 K.\cite{bayrakci04,sales04} The powder
shows no magnetic transition above 2 K and the first order
transition at higher temperatures is smeared over the temperature
range from 250 to 310 K as a series of first order transitions.
The different properties between powder and single crystal samples
is likely related to the real Na amount in the structure of
Na$_x$CoO$_2$ and the distribution of Na in Na(1) and Na(2) sites.
We would like to emphasize that despite the evidence for A-type
antiferromagnetic correlations in $x\geq 0.75$
samples\cite{0410224} a clear demonstration of the
antiferromagnetism by means of a static magnetic order (elastic
magnetic Bragg peaks) is still lacking. Possibly, this is the
cause for the broad magnetically split M\"{o}ssbauer spectra below
$T\sim 7 K$.

Summarizing we studied the Na$_{0.82}$CoO$_2$ (doped with 1\%
$^{57}$Fe) using M\"{o}ssbauer and NMR spectroscopies, dc
magnetization measurements and x-ray diffraction data. The
M\"{o}ssbauer data clearly show that the deficient charge
reservoir of Na-planes creates a mixed valance compound comprising
from, Co$^{3+}$ and Co$^{4+}$. Parts of the M\"{o}ssbauer spectrum
corresponding to Fe$^{4+}$ and Fe$^{3+}$ becomes magnetically
split below 10 K. The magnetic spectra may imply a magnetic glass
phase or slow electronic spin relaxation phenomena, because the
magnetization measurements do not show any characteristic feature
corresponding to a magnetic transition. The different quadrupole
splitting of the two iron valence states indicate a charge ordered
structure.

\end{document}